\documentclass[sigconf, nonacm]{acmart}

\renewcommand\footnotetextcopyrightpermission[1]{}
\usepackage{booktabs}
\usepackage{xcolor}
\usepackage{colortbl}
\usepackage{array}
\usepackage{placeins}

\definecolor{acblue}{HTML}{2171B5}

\newcommand{\feat}[1]{\texttt{#1}}

\hypersetup{
  colorlinks=true,
  linkcolor=acblue,
  citecolor=acblue,
  urlcolor=acblue,
}

\begin{document}

\title{Predicting Activity Cliffs for Autonomous Medicinal Chemistry}

\author{Michael F. Cuccarese, PhD}
\affiliation{%
  \country{}}
\email{}

\begin{abstract}
Activity cliff prediction -- identifying positions where small structural changes cause large potency shifts -- has been a persistent challenge in computational medicinal chemistry. This work focuses on a parsimonious definition: which small modifications, at which positions, confer the highest probability of an outcome change. Position-level sensitivity is calculated using 25~million matched molecular pairs from 50~ChEMBL targets across six protein families, revealing that two questions have fundamentally different answers. ``Which positions vary most?'' is answered by scaffold size alone (NDCG@3 = 0.966), requiring no machine learning. ``Which are true activity cliffs?'' -- where \textit{small} modifications cause \textit{disproportionately large} effects, as captured by SALI normalization -- requires an 11-feature model with 3D pharmacophore context (NDCG@3 = 0.910 vs.\ 0.839 random), generalizing across all six protein families, novel scaffolds (0.913), and temporal splits (0.878). The model identifies the cliff-prone position first 53\% of the time (vs.\ 27\% random -- 2$\times$ lift), reducing positions a chemist must explore from 3.1 to 2.1 -- a 31\% reduction in first-round experiments. Predicting \textit{which} modification to make is not tractable from structure alone (Spearman 0.268, collapsing to $-$0.31 on novel scaffolds). The system is released as open-source code and an interactive webapp.
\end{abstract}

\maketitle

\section{Introduction}

Drug discovery campaigns are not limited by the number of compounds chemists can synthesize. They are limited by \textit{which} compounds get made. Combinatorial chemistry and parallel synthesis platforms transformed throughput in the 1990s, yet as Brown and Bostr\"{o}m documented across 25~years of medicinal chemistry literature, the reaction toolkit used in practice has barely changed: the same amide couplings, reductive aminations, and Suzuki couplings dominate lead optimization today as they did in 2000~\cite{Brown2016}. The problem is not synthetic capacity; it is that high-throughput approaches are constrained to the same chemistries, and so explore the same corners of chemical space.

A smarter approach would target synthesis effort precisely -- make fewer compounds, but make the ones that carry the most information about the structure--activity relationship (SAR). Depending on the chemical throughput of the organization, even a modest efficiency gain through avoided low-impact compounds would be of considerable benefit. This is also in alignment with the premise of active learning in drug discovery~\cite{Reker2015,Reker2020}: select the experiment that maximally reduces uncertainty about SAR, rather than the most synthetically convenient experiment or the compound with the highest predicted potency.

Activity cliff probes are the highest-value experiments in an information-poor setting, where existing results haven't appreciably fed into local models. An activity cliff is a position on a molecule where a small structural change causes a large change in biological activity~\cite{Guha2008,Maggiora2006}. Identifying such positions before synthesis -- from structure alone -- would allow a computational system to propose two or three maximally informative modifications, generate specific synthesizable compounds that realize them, and hand a ranked list to a chemist or automated synthesis platform.

This paper asks: \textbf{can we predict activity cliff positions well enough to guide autonomous first-round SAR exploration?}

The answer depends critically on how the question is framed. Under raw sensitivity metrics, a trivial heuristic -- rank positions by inverse scaffold size (i.e., positions on smaller scaffolds ranked first) -- appears to solve the problem (NDCG@3 = 0.966) with no machine learning required. This is not a bug; it validates at massive scale a well-known chemical principle: R-group contributions to binding scale inversely with scaffold size~\cite{Hopkins2004}. But, without rewarding smallness of change, it is weighted toward existing and obvious ways to break a compound's activity as opposed to those which have impact while still preserving much of the scaffold's character.

SALI normalization~\cite{Guha2008} -- dividing activity change by structural change -- isolates positions where \textit{small} modifications cause \textit{disproportionately large} effects. Under SALI, the scaffold-size heuristic falls below the random baseline (0.791 vs.\ 0.839), while an 11-feature model incorporating 3D pharmacophore context achieves 0.910 -- capturing 44\% of available headroom over chance. Machine learning genuinely adds value, but only when the right question is asked.

\paragraph{Contributions.} (1)~A position-level reframing of activity cliff prediction using 25~million MMPs across 50~targets, with the largest systematic validation to date. (2)~A demonstration that raw MMP-derived sensitivity conflates position vulnerability with modification size, reversing method rankings under SALI normalization -- a cautionary finding for prior work. (3)~An 11-feature, target-agnostic model that generalizes across protein families by encoding the pharmacophore as the local context as opposed to the target name, novel scaffolds, and temporal splits. (4)~An honest negative result: predicting \textit{which} type of modification to make at each position is not tractable from structure alone. (5)~A deployed webapp and open-source tool providing end-to-end compound recommendations from a single SMILES input.

\section{Related Work}

\paragraph{Activity cliff prediction.} Maggiora~\cite{Maggiora2006} formalized the concept of activity cliffs as discontinuities in the structure--activity landscape. Stumpfe and Bajorath~\cite{Stumpfe2012} systematically catalogued activity cliffs across public databases, establishing the prevalence of the phenomenon. Van Tilborg et al.~\cite{vanTilborg2022} trained neural models to predict activity cliffs but demonstrated that performance degrades severely out-of-distribution -- the fundamental limitation that motivates the target-agnostic approach taken here.

\paragraph{Matched molecular pairs.} The Hussain--Rea algorithm~\cite{Hussain2010} and subsequent implementations~\cite{Dalke2018} enable exhaustive enumeration of structurally matched compound pairs from activity databases. MMPs have become a standard tool for SAR analysis, but prior work has primarily used them descriptively (cataloguing known cliffs) rather than predictively (identifying cliffs before measurement).

\paragraph{SALI.} The Structure--Activity Landscape Index, introduced by Guha and Van Drie~\cite{Guha2008}, normalizes activity change by structural similarity to isolate true activity cliffs from trivial size-dependent effects. Despite its importance, SALI normalization has been inconsistently applied in the activity cliff prediction literature, leading to inflated performance claims that this work explicitly addresses.

\paragraph{Active learning in drug discovery.} Reker et al.~\cite{Reker2015,Reker2020} demonstrated that Bayesian active learning can reduce the number of compounds needed to map SAR. Bellamy et al.~\cite{Bellamy2022} extended this with batched experimental design. The present system is complementary: it addresses the \textit{first round} of experimentation, before any target-specific activity data exist, where active learning cannot yet operate.

\paragraph{Ligand efficiency.} The principle that smaller molecules make more efficient use of their binding interactions~\cite{Hopkins2004} provides the physical basis for the scaffold-size heuristic observed here. Our results validate this principle at unprecedented scale (25M MMPs, 50 targets) while showing that it is necessary but not sufficient for identifying true activity cliffs.

\section{Methods}

\subsection{MMP Extraction and Position Sensitivity}

Single-cut MMPs were extracted from ChEMBL~36 (2024 release) using the Hussain--Rea algorithm~\cite{Hussain2010} via RDKit's \texttt{rdMMPA}, yielding 25~million pairs across 50~targets spanning six protein families (22~kinases, 7~enzymes, 7~immune targets, 4~epigenetic targets, 3~ion channels, 7~receptors/other).

For each substitution position (attachment point on a molecular scaffold) per target, two sensitivity measures were computed:

For each position, \textbf{raw sensitivity} is the mean~$|\Delta p\mathrm{Activity}|$ across all MMPs. \textbf{SALI sensitivity}~\cite{Guha2008} normalizes by modification size:
\begin{equation*}
  \text{mean}\!\left(\frac{|\Delta p\mathrm{Activity}|}
       {\max(\text{R-group heavies}) + 1}\right)
\end{equation*}

This produced 598{,}173 position-level training examples across 133{,}763 molecules (median 4 positions per molecule).

\subsection{Feature Architecture}

The model uses 11 structural features computed from the molecule alone, with no activity data required at prediction time (Table~\ref{tab:features}).

\begin{table}[ht]
\caption{Feature architecture: two topological and nine 3D pharmacophore context features (all computed within 4~\AA{} of the attachment point).}
\label{tab:features}
\small
\begin{tabular}{@{}p{1.3in}l@{}}
\toprule
\textbf{Topology (2)} \\
\midrule
\feat{core\_n\_heavy}   & Heavy atom count \\
\feat{core\_n\_rings}    & Ring count \\
\addlinespace
\textbf{3D context (9)} & \textit{within 4~\AA{} of attach.\ point} \\
\midrule
\feat{n\_donor}       & H-bond donors \\
\feat{n\_acceptor}    & H-bond acceptors \\
\feat{n\_hydrophobic} & Hydrophobic atoms \\
\feat{n\_aromatic}    & Aromatic atoms \\
\feat{sasa}           & Solvent-accessible SA \\
\feat{gasteiger}      & Partial charge \\
\feat{n\_rotbonds}    & Rotatable bonds \\
\feat{is\_aromatic}   & Aromatic attachment \\
\feat{n\_heavy}       & Heavy atom density \\
\bottomrule
\end{tabular}
\end{table}

The topology features capture scaffold complexity. The 3D context features describe the local chemical environment surrounding each substitution position, enabling the model to distinguish pharmacophore-critical positions (near H-bond networks, aromatic stacking surfaces) from solvent-exposed, tolerant positions.

\subsection{Model and Validation}

HistGradientBoosting (HGB) regression models were trained with leave-one-target-out cross-validation across all 50~targets. Primary metric: NDCG@3 (ranking quality of the top-3 positions per molecule). Secondary metrics: Spearman~$\rho$, Hit@1 (fraction of molecules where the most sensitive position is ranked first).

Statistical tests: paired Wilcoxon signed-rank on per-target NDCG@3, bootstrap 95\% confidence intervals (10{,}000 resamples, seed~=~42), Cohen's~$d$ for effect sizes, 1{,}000-permutation null distributions.

\subsection{Out-of-Distribution Validation Strategy}

The central concern for any predictive model in chemistry is information leakage: if training and test compounds share scaffolds, chemical series, or target-specific SAR patterns, apparent generalization is illusory. We designed 11~OOD stress tests to address this systematically. Three matter most, listed in order of increasing stringency:

Temporal holdout (train on data published before a cutoff, test on data published after) is the conventional gold standard, but it provides weaker protection than commonly assumed. Chemical series routinely span a decade of publications, so temporal splits still share scaffold chemotypes across the boundary. Novel scaffold holdout -- withholding the 20\% most structurally dissimilar scaffolds -- directly tests whether predictions transfer to chemistry the model has never seen, regardless of when it was published. External validation on datasets entirely outside ChEMBL (COVID Moonshot, Open Force Field, Schr\"{o}dinger FEP) is the most stringent test of all: different targets, different laboratories, different chemical matter, with no possibility of information leakage from the training corpus.

Additional OOD experiments include: feature ablation (11~model configurations), target family holdout, feature robustness (conformer sensitivity), learning curve (5--100\% training data), directional analysis, permutation tests (1{,}000 label shuffles), calibration, and hyperparameter sensitivity (50~random configurations).

\section{Calibration: Raw Sensitivity Is Dominated by Scaffold Size}

Before presenting the main results, a calibration. Under raw sensitivity (mean~$|\Delta p\mathrm{Activity}|$, unnormalized), a trivial heuristic -- rank positions by inverse scaffold size -- achieves NDCG@3 = 0.966, matching or exceeding the full 11-feature ML model (0.959) on 48 of 50~targets. The physical interpretation is straightforward: on a 10-atom scaffold, swapping methyl for cyclohexyl changes the molecule by ${\sim}$50\%; on a 28-atom scaffold, the same swap is ${\sim}$20\%. R-group contributions to binding scale inversely with scaffold size~\cite{Hopkins2004}.

This validates ligand efficiency theory at scale, but it does not identify activity cliffs. The heuristic excels because positions on smaller scaffolds host \textit{larger} R-group modifications in ChEMBL, and larger modifications trivially produce larger activity changes. Raw sensitivity conflates genuine position vulnerability with modification size -- a confound that SALI normalization was designed to resolve.

\section{SALI Normalization: Identifying True Activity Cliffs}

The Structure--Activity Landscape Index~\cite{Guha2008} was designed precisely to address the confound in raw sensitivity. SALI normalizes activity change by structural change:
\begin{equation}
  \text{SALI sensitivity} = \text{mean}\!\left(
    \frac{|\Delta p\mathrm{Activity}|}
         {\max(\text{R-group heavy atoms}) + 1}
  \right)
\end{equation}
This is a position-level adaptation of the original SALI~\cite{Guha2008}, which normalizes by Tanimoto dissimilarity between compound pairs. At the MMP level, structural similarity is dominated by the shared core; the meaningful variation is the R-group size, which heavy atom count captures directly. This adaptation isolates positions where small modifications cause disproportionately large activity shifts -- the true activity cliffs that carry the most information per experiment.

\subsection{The Model Separates from Chance}

Under SALI normalization, random baseline NDCG@3 is 0.839 -- the performance of a model with no information. The 11-feature ML model achieves 0.910, capturing 44\% of the available headroom between chance and a perfect ranker (Table~\ref{tab:sali}). The scaffold-size heuristic, which dominated raw metrics, collapses to 0.791 -- \textit{below random} -- because SALI strips away the very confound the heuristic exploits.

\begin{table}[ht]
\centering
\caption{Model performance under SALI normalization. The ML model separates meaningfully from random; the heuristic falls below chance.}
\label{tab:sali}
\small
\begin{tabular}{@{}lcc@{}}
\toprule
Model & SALI NDCG@3 & vs.\ random \\
\midrule
Random baseline          & 0.839 & -- \\
Global mean              & 0.831 & $-$0.008 \\
Heuristic (scaffold size) & 0.791 & $\mathbf{-0.048}$ \\
3D context only (9 feat) & 0.875 & $+$0.036 \\
Topology only (2 feat)   & 0.909 & $+$0.070 \\
\textbf{Full model (11 feat)} & \textbf{0.910} & $\mathbf{+0.071}$ \\
\bottomrule
\end{tabular}
\end{table}

\paragraph{Why the scores look high -- and what they mean in practice.} The typical molecule in this dataset has 4 modifiable positions (median; mean 5.4; 84\% have 3--6). With so few items to rank, even a random shuffle scores ${\sim}$0.84 NDCG@3 -- there simply aren't many ways to be wrong. The meaningful signal is the lift above random.

A more intuitive metric is Hit@1: the fraction of molecules where the model correctly identifies the most cliff-prone position on its first try. Random guessing gets this right 26.6\% of the time. The SALI model gets it right 52.7\% of the time -- a 2$\times$ lift. By the second position tested, the model has found the true cliff 74.1\% of the time (vs.\ 49.2\% random).

In terms of experimental efficiency: a medicinal chemist guided by this model needs to test on average 2.1~positions to find the most cliff-prone site, versus 3.1~positions by random exploration -- saving roughly one position per molecule, a 31\% reduction in experiments. Across a 10-scaffold campaign at ${\sim}$10 compounds per position, that translates to roughly 100 fewer compounds to learn the same SAR.

The Spearman correlation between scaffold size and ground-truth position sensitivity flips from $r = -0.506$ (raw: smaller scaffolds = higher sensitivity) to $r = +0.301$ (SALI: smaller scaffolds = \textit{lower} cliff propensity once modification size is controlled). Under SALI, the ``obvious'' answer is not merely uninformative -- it is actively misleading, ranking worse than guessing. The 11-feature model resists the collapse because its 3D pharmacophore context features capture information orthogonal to scaffold size: whether a position sits near H-bond networks, aromatic surfaces, or solvent-exposed regions.

\subsection{Cross-Family Generalization}

\begin{figure*}[t]
  \centering
  \includegraphics[width=\textwidth]{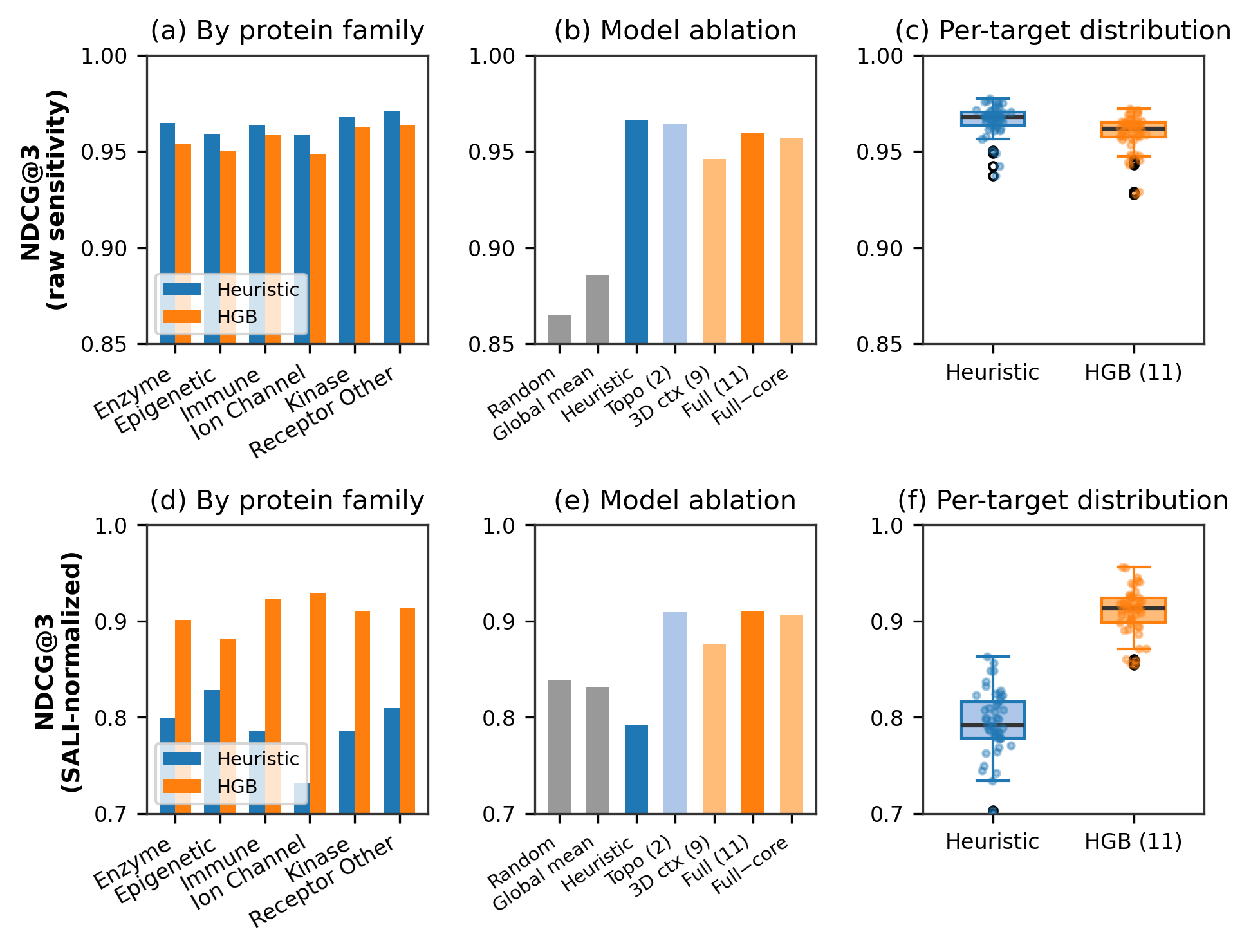}
  \caption{Raw vs.\ SALI-normalized position sensitivity across six protein families and model ablations. \textbf{Top row} (raw sensitivity): the heuristic and ML model perform comparably across all six families. \textbf{Bottom row} (SALI-normalized): the heuristic collapses across all families while the ML model holds, with a consistent advantage over random (0.839) in every family.}
  \label{fig:universality}
\end{figure*}

The separation from random is not driven by a few well-behaved targets. Leave-one-target-out NDCG@3, grouped by protein family, shows the ML model holds above random across all six families (Figure~\ref{fig:universality}, Table~\ref{tab:family}).

\begin{table}[ht]
\centering
\caption{Per-family SALI NDCG@3 (leave-one-target-out, grouped by family). The ML model holds above random in every family.}
\label{tab:family}
\small
\begin{tabular}{@{}lccc@{}}
\toprule
Family & $N$ & NDCG@3 & vs.\ rand. \\
\midrule
Enzyme          & 7  & 0.901 & $+$0.062 \\
Epigenetic      & 4  & 0.881 & $+$0.042 \\
Immune          & 7  & 0.923 & $+$0.084 \\
Ion channel     & 3  & 0.929 & $+$0.090 \\
Kinase          & 22 & 0.911 & $+$0.072 \\
Receptor/other  & 7  & 0.913 & $+$0.074 \\
\bottomrule
\end{tabular}
\end{table}

The range is tight: 0.881--0.929 across families with very different binding site architectures. The model is target-agnostic in practice, not just in principle.

\subsection{Two Questions, Two Answers}

These results establish that the question determines the answer:

\begin{table}[ht]
\centering
\small
\begin{tabular}{@{}p{1.6in}lc@{}}
\toprule
Question & Best predictor & NDCG@3 \\
\midrule
\raggedright Which positions vary most? & Heuristic & 0.966 \\
\raggedright Which are true cliffs?     & HGB (11~feat) & 0.910 \\
\bottomrule
\end{tabular}
\end{table}

\begin{figure*}[t]
  \centering
  \includegraphics[width=\textwidth]{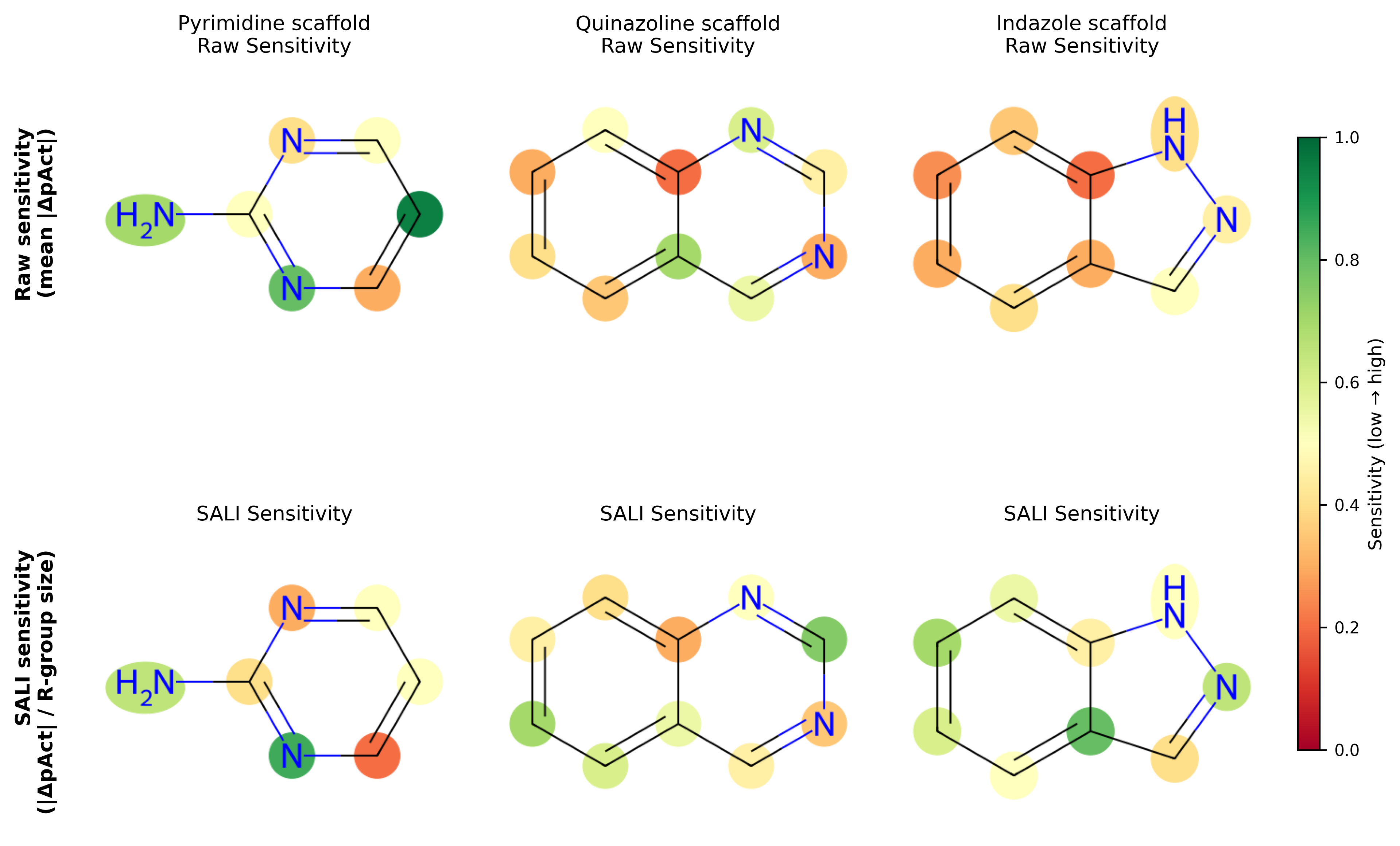}
  \caption{Raw vs.\ SALI sensitivity rankings on representative molecules. Atoms colored by predicted sensitivity (green = low, red = high). The highest-ranked position differs between raw and SALI across all scaffolds, reflecting the sign-flip in scaffold-size correlation ($r = -0.506 \to +0.301$).}
  \label{fig:structure}
\end{figure*}

Both questions are useful in different contexts. The heuristic quickly identifies where SAR \textit{lives} on the molecule. The SALI model identifies where \textit{small changes cause large effects} -- the highest-value positions for tight optimization. The two rankings disagree on the same molecules (Figure~\ref{fig:structure}), with the sign of the scaffold-size correlation flipping from $-0.506$ to $+0.301$.

\section{From Cliff Identification to Compound Selection}

\begin{figure*}[t]
  \centering
  \includegraphics[width=\textwidth]{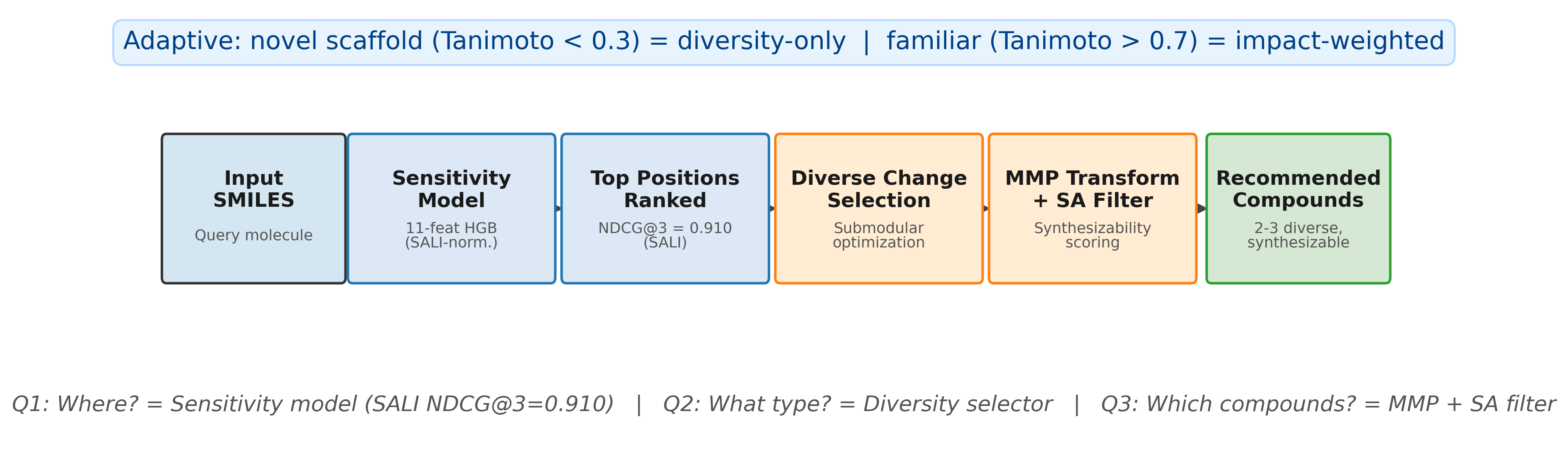}
  \caption{Complete pipeline from input SMILES to synthesizable compound recommendations. The SALI-normalized sensitivity model ranks positions; submodular optimization selects diverse change types; MMP transforms from the 25M-pair corpus are retrieved, filtered for synthesizability, and output as a ranked compound list.}
  \label{fig:pipeline}
\end{figure*}

Identifying \textit{where} activity cliffs occur is the core contribution. But a practical system must also recommend \textit{what to make} at those positions (Figure~\ref{fig:pipeline}). Two strategies are natural: predict which modification type will have the largest impact (an \textbf{impact-weighted} approach), or cover the space of possible modification types as broadly as possible (a \textbf{diversity-weighted} approach).

\subsection{Change-Type Prediction: Why Impact Weighting Fails}

We attempted to predict \textit{which} type of modification (e.g., electron-withdrawing, H-bond donor, size increase) produces the largest activity shift at each position. If this worked, a system could recommend specific compounds rather than just positional priorities.

The change-type model ranks 11 R-group property modification axes by predicted impact at each position. Its performance is measured by Spearman~$\rho$ -- the rank correlation between predicted and observed impact across the 11 axes. A Spearman of 1.0 would mean the model perfectly ranks which modification types matter most; 0.0 would mean no predictive signal at all.

\begin{figure}[ht]
  \centering
  \includegraphics[width=\columnwidth]{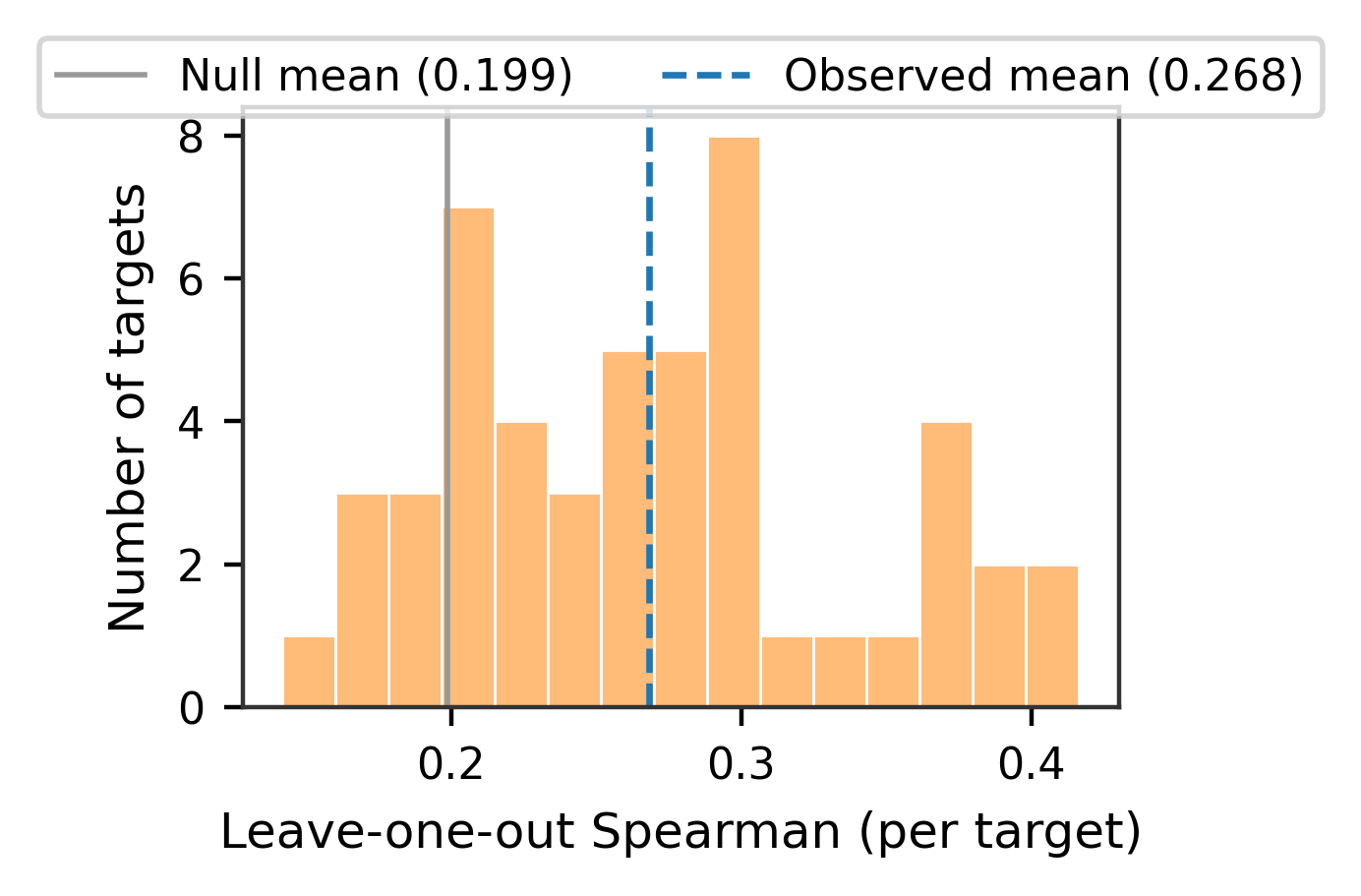}
  \caption{Change-type prediction across 50~targets. The observed mean Spearman (0.268, dashed) exceeds the permutation null (0.199) but explains only ${\sim}7$\% of variance -- insufficient for actionable compound selection.}
  \label{fig:changetype}
\end{figure}

Mean Spearman~$\rho = 0.268$ across 50~targets (Figure~\ref{fig:changetype}). A permutation null -- the same model trained on randomly shuffled labels -- achieves 0.199. So the signal is real (the model has learned \textit{something} about which modifications matter), but it explains only ${\sim}7$\% of variance. In practical terms, the top-ranked modification type is correct only ${\sim}56$\% of the time.

More critically, this weak in-sample signal collapses on novel scaffolds ($\rho = -0.31$), meaning the model actively predicts the \textit{wrong} modification types when deployed on new chemistry. SALI normalization worsens performance further ($\rho = 0.192$), confirming that even the in-sample signal reflects modification size rather than genuine pharmacophore insight.

\paragraph{Implication.} Predicting \textit{where} on a molecule to modify is tractable from structure alone; predicting \textit{what specific change to make} is not. Crossing this boundary will require target-specific activity data, which is unavailable at campaign start.

\subsection{Diversity-Weighted Selection: Covering the Hypothesis Space}

Given that impact prediction fails on novel chemistry, the alternative is information-theoretic: instead of betting on which modification type will matter most, cover the space of possible modifications as broadly as possible. The logic is straightforward -- if we cannot predict which of the 11 modification axes will produce a cliff, we should ensure our first experiments sample across as many axes as possible, maximizing the probability that at least one experiment hits the true SAR driver.

At each sensitive position, greedy submodular optimization selects 2--3 modifications that maximally cover the SAR hypothesis space across 11~change-type axes (electron-withdrawing/donating, H-bond donor/acceptor, lipophilicity, size, aromaticity, sp3~character, ring count). Synthesizability filtering uses MMP transforms from the 25M-pair corpus with SA~score thresholds~\cite{Swanson2024,Gao2020}, ensuring compounds are concretely makeable.

\subsection{The Complete Pipeline}

Given only a scaffold SMILES, the system: (1)~identifies true activity cliff positions via the SALI-normalized model; (2)~selects maximally diverse chemical hypotheses at each position; (3)~retrieves synthesizable MMP transforms from the 25M-pair corpus; and (4)~outputs a ranked compound list. This reduces the typical 20--40~compound first round to 6--9 targeted experiments without sacrificing SAR coverage.

\section{Generalization: Does It Work on New Chemistry?}

The results in Sections~5--6 establish that the model works in-distribution. This section consolidates all out-of-distribution evidence -- for both cliff identification (SALI NDCG@3) and compound selection -- to address the question that matters most: will it work on chemistry the model has never seen?

\subsection{Novel Scaffold Holdout}

The most direct test of structural generalization. When 20\% of the most structurally dissimilar scaffolds (by Tanimoto distance to all training cores) are held out, the cliff identification model maintains SALI NDCG@3 = 0.913 -- degrading only 0.003 from the LOO baseline (0.910). The random baseline under the same split is 0.839. The model's separation from chance holds on chemistry it has structurally never seen.

\subsection{Temporal Split}

Under the strictest temporal split (train~$\leq$~2015, test~$>$~2015), 88\% of test scaffolds are novel and 83\% of test cores are novel. The cliff identification model maintains SALI NDCG@3 = 0.878 -- a modest but consistent advantage over random (0.839) that holds across both temporal cutoffs tested.

\subsection{External Validation}

The ultimate test: 114~positions from three datasets entirely outside ChEMBL, with 67.8\% novel scaffolds and 94.7\% novel compounds. The diversity-weighted compound selector achieves a top-hit rate (fraction of positions where the highest-impact change axis is among the selector's top-3 recommendations) of 0.439 vs.\ 0.271 for random selection (Wilcoxon $p = 1.86 \times 10^{-5}$). The impact-weighted approach underperforms random (0.193), confirming the in-sample negative result: when change-type prediction is unreliable, \textit{covering the hypothesis space} outperforms \textit{betting on a specific prediction}.

\begin{table}[ht]
\centering
\caption{External validation (114~positions from COVID Moonshot, Open Force Field, Schr\"{o}dinger FEP). Top-hit rate = fraction of positions where the highest-impact change axis is among the selector's top-3 recommendations.}
\label{tab:external}
\small
\begin{tabular}{@{}lcc@{}}
\toprule
Strategy & Top-hit rate & vs.\ random \\
\midrule
Random              & 0.271 & -- \\
Impact-weighted     & 0.193 & $-$0.078 \\
\textbf{Diversity-weighted}  & \textbf{0.439} & $\mathbf{+0.168}$ \\
\bottomrule
\end{tabular}
\end{table}

\subsection{The Signal Is Structural, Not Statistical}

Two additional experiments (reported under raw sensitivity, where the model is most constrained) confirm that the model captures physical relationships rather than statistical artifacts:

\paragraph{Learning curve.} Raw NDCG@3 plateaus at 5\% of training data (0.958 vs.\ 0.959 at 100\%). Even with only 50~positions per target (simulating a rare target), NDCG@3 = 0.949. The model learns general relationships between pharmacophore context and position sensitivity, not target-specific patterns.

\paragraph{Hyperparameter sensitivity.} Fifty random HGB configurations produce NDCG@3 values spanning only 0.954--0.965 (range = 0.011). The performance ceiling is inherent to the feature set, not a tuning artifact.

\subsection{Sensitivity Does Not Predict Direction}

Does high sensitivity mean a position is ``fragile'' (modifications hurt potency) or ``improvable'' (modifications can increase potency)? The answer is neither. High-sensitivity positions split almost exactly: 29.2\% improvable, 41.7\% neutral, 29.0\% fragile. The correlation between sensitivity and directionality is Spearman~$= -0.005$ ($p = 0.19$). ``Sensitive'' means \textit{variable} -- the position matters, but which direction depends on the specific modification and target, not on the structural context the model can observe.

\section{Discussion}

\subsection{What Works}

\textbf{Position sensitivity prediction is solved for its intended purpose.} SALI-normalized NDCG@3 = 0.910 (vs.\ 0.839 random) provides reliable, target-agnostic identification of true activity cliff positions from structure alone. The model identifies the most cliff-prone position on its first try 53\% of the time (vs.\ 27\% random), reducing the average positions a chemist must explore from 3.1 to 2.1 -- a 31\% reduction in first-round experiments. It generalizes across protein families, novel scaffold holdouts (0.913), temporal splits (0.878), and external datasets entirely outside the training corpus.

\textbf{The SALI reframe is methodologically important.} Raw MMP-derived sensitivity conflates position vulnerability with modification size; SALI normalization reverses relative method rankings. This is a cautionary finding for prior work~\cite{Stumpfe2012,vanTilborg2022} that evaluated activity cliff prediction under raw metrics: results that appeared strong may have been measuring the obvious (scaffold size) rather than the interesting (pharmacophore context).

\subsection{The ``Where'' vs.\ ``What'' Boundary}

This work establishes a clear boundary: \textit{where} to modify a molecule is predictable from structure alone; \textit{what specific change to make} is not. The change-type model's in-sample $\rho = 0.268$ collapses to $\rho = -0.31$ on novel scaffolds -- historical activity-change patterns from ChEMBL do not transfer to unseen chemistry~\cite{Tyrchan2017}. Crossing this boundary will require target-specific activity data, which is unavailable at campaign start. Until then, diversity-weighted selection -- covering the hypothesis space rather than betting on a single prediction -- is the rational strategy, and external validation confirms this ($+$0.168 top-hit rate over random).

\subsection{Path Forward}

The right paradigm is \textbf{closed-loop experimental feedback}. This system addresses the first round: identify sensitive positions, propose diverse modifications, and collect data. Accumulated measurements can then enable target-specific change-type predictions via active learning~\cite{Reker2020,vanTilborg2022} and batched Bayesian optimization~\cite{Bellamy2022}.

\section{Limitations}

\begin{enumerate}
  \item \textbf{Single-cut MMPs assume position independence.} Cooperative SAR -- where modifications at two positions interact -- is invisible to single-cut MMP analysis. Multi-cut MMP methods~\cite{Leach2006} could extend the approach but increase computational cost substantially.

  \item \textbf{Sensitivity does not predict direction.} The model identifies \textit{where} activity cliffs occur but cannot predict whether modification will improve or degrade potency. This requires protein structure information not captured by ligand-only features.

  \item \textbf{Free-ligand 3D features are approximate.} Conformers are generated for the free ligand, not the protein-bound state. Gasteiger charges are crude compared to quantum-mechanical calculations. These approximations are pragmatic -- they enable zero-install predictions from SMILES alone -- but bound the model's resolution.

  \item \textbf{Target selection bias.} The 50~ChEMBL targets were chosen for data richness (thousands of MMPs each). Performance on data-poor targets or novel target classes is not directly validated, though the learning curve analysis suggests the model is relatively data-efficient.

  \item \textbf{Retrospective validation only.} No prospective experimental validation was performed. The external validation on COVID Moonshot, Open Force Field, and Schr\"{o}dinger FEP provides the strongest available test of generalization, but ultimate validation requires synthesis and measurement.

  \item \textbf{Change-type prediction explains only ${\sim}7$\% of variance.} The ceiling for ligand-only change-type prediction appears to be low. Protein-side features (binding pocket shape, interaction fingerprints) would likely improve this, but are unavailable in the target-agnostic setting.
\end{enumerate}

\section{Interactive Demo}

A prediction model is only useful if practitioners can apply it without friction. The system is deployed as an interactive Streamlit webapp where users can: (1)~enter any SMILES string; (2)~see positions colored by predicted SALI sensitivity; (3)~browse recommended modifications with diversity scores; (4)~look up supporting evidence from the 25M MMP corpus; and (5)~export ranked compound lists.

The webapp includes honest framing: position sensitivity is dominated by scaffold size under raw metrics; the SALI model adds value for identifying true cliffs; change-type recommendations are weak and should be treated as hypotheses, not predictions. Flexibility warnings flag molecules with high rotatable bond counts where conformer-dependent features are less reliable.

For integration into autonomous pipelines, the core library is released as open-source Python code. Given a SMILES, \texttt{predict.py} returns position-level SALI sensitivity scores and ranked compound recommendations in a single function call.

\section{Conclusion}

Activity cliff prediction has been framed as a machine learning problem requiring target-specific training data. This work shows that the problem has two distinct components with different solutions. Position sensitivity ranking -- \textit{where} on a molecule SAR lives -- is dominated by scaffold size, a chemical principle that requires no learning. True activity cliff identification -- \textit{where small changes cause disproportionately large effects} -- requires 3D pharmacophore context features and machine learning, achieving NDCG@3 = 0.910 target-agnostically across 50~targets and six protein families, with robust generalization to novel scaffolds (0.913) and future chemistry (0.878).

The practical system reduces a first-round SAR campaign from 20--40 unfocused compounds to 6--9 targeted experiments at predicted cliff positions. By identifying the cliff-prone position on its first try 53\% of the time -- versus 27\% by chance -- the model saves roughly one analog series per scaffold (2.1 vs.\ 3.1 positions to explore), or ${\sim}$100 fewer compounds across a typical 10-scaffold campaign. The underlying insight is one of right-sizing the question: asking a model \textit{where} activity cliffs live is tractable; asking it \textit{what specific change to make} is not. The system's most important limitation defines the natural handoff point to closed-loop experimental feedback and active learning.

\paragraph{Code and data availability.} The complete system, including trained models, evaluation scripts, and the interactive webapp, is available at \url{https://github.com/mcuccarese/activity-cliffs}. Activity data were extracted from ChEMBL~36, a publicly available database~\cite{Zdrazil2024}.

\paragraph{Acknowledgments.} Claude (Anthropic) was used for code development, data analysis, and manuscript preparation assistance.

\bibliographystyle{ACM-Reference-Format}

\begin{thebibliography}{16}

\bibitem{Brown2016}
D.~G. Brown and J. Bostr\"{o}m.
\newblock Analysis of past and present synthetic methodologies on medicinal chemistry.
\newblock \textit{J. Med. Chem.}, 59:4443--4458, 2016.

\bibitem{Reker2015}
D.~Reker and G.~Schneider.
\newblock Active-learning strategies in computer-assisted drug discovery.
\newblock \textit{Drug Discov. Today}, 20:458--465, 2015.

\bibitem{Reker2020}
D.~Reker.
\newblock Practical considerations for active machine learning in drug discovery.
\newblock \textit{Drug Discov. Today: Technol.}, 32--33:73--79, 2020.

\bibitem{Guha2008}
R.~Guha and J.~H. Van~Drie.
\newblock Structure--activity landscape index: identifying and quantifying activity cliffs.
\newblock \textit{J. Chem. Inf. Model.}, 48:646--658, 2008.

\bibitem{Maggiora2006}
G.~M. Maggiora.
\newblock On outliers and activity cliffs -- why QSAR often disappoints.
\newblock \textit{J. Chem. Inf. Model.}, 46:1535, 2006.

\bibitem{Hopkins2004}
A.~L. Hopkins, C.~R. Groom, and A.~Alex.
\newblock Ligand efficiency: a useful metric for lead selection.
\newblock \textit{Drug Discov. Today}, 9:430--431, 2004.

\bibitem{Stumpfe2012}
D.~Stumpfe and J.~Bajorath.
\newblock Exploring activity cliffs in medicinal chemistry.
\newblock \textit{J. Med. Chem.}, 55:2932--2942, 2012.

\bibitem{vanTilborg2022}
D.~van Tilborg, A.~Alenicheva, and F.~Grisoni.
\newblock Exposing the limitations of molecular machine learning with activity cliffs.
\newblock \textit{J. Chem. Inf. Model.}, 62:5938--5951, 2022.

\bibitem{Hussain2010}
J.~Hussain and C.~Rea.
\newblock Computationally efficient algorithm to identify matched molecular pairs.
\newblock \textit{J. Chem. Inf. Model.}, 50:339--348, 2010.

\bibitem{Dalke2018}
A.~Dalke, J.~Hert, and C.~Kramer.
\newblock mmpdb: An open-source matched molecular pair platform for large multiproperty data sets.
\newblock \textit{J. Chem. Inf. Model.}, 58:902--910, 2018.

\bibitem{Bellamy2022}
H.~Bellamy, A.~Abdel~Rehim, O.~I. Orhobor, and R.~King.
\newblock Batched Bayesian optimization for drug design in noisy environments.
\newblock \textit{J. Chem. Inf. Model.}, 62:3970--3981, 2022.

\bibitem{Swanson2024}
K.~Swanson, G.~Liu, D.~B. Catacutan, A.~Arnold, J.~Zou, and J.~M. Stokes.
\newblock Generative AI for designing and validating easily synthesizable and structurally novel antibiotics.
\newblock \textit{Nat. Mach. Intell.}, 6:338--353, 2024.

\bibitem{Gao2020}
W.~Gao and C.~W. Coley.
\newblock The synthesizability of molecules proposed by generative models.
\newblock \textit{J. Chem. Inf. Model.}, 60:5714--5723, 2020.

\bibitem{Tyrchan2017}
C.~Tyrchan and E.~Evertsson.
\newblock Matched molecular pair analysis in short: algorithms, applications and limitations.
\newblock \textit{Comput. Struct. Biotechnol. J.}, 15:86--90, 2017.

\bibitem{Leach2006}
A.~G. Leach, H.~D. Jones, D.~A. Cosgrove, P.~W. Kenny, L.~Ruston, P.~MacFaul, J.~M. Wood, N.~Colclough, and B.~Law.
\newblock Matched molecular pairs as a guide in the optimization of pharmaceutical properties; a study of aqueous solubility, plasma protein binding and oral exposure.
\newblock \textit{J. Med. Chem.}, 49:6672--6682, 2006.

\bibitem{Zdrazil2024}
B.~Zdrazil, E.~Felix, F.~Hunter, E.~J. Manners, J.~Blackshaw, E.~M. Sheridan, A.~R. Leach, et~al.
\newblock The ChEMBL Database in 2023: a drug discovery platform spanning genomics, bioactivity and beyond.
\newblock \textit{Nucleic Acids Res.}, 52:D1180--D1192, 2024.

\end{thebibliography}

\end{document}